\documentclass[10pt,conference]{IEEEtran}

\usepackage{graphicx}
\graphicspath{ {figures/}}
\usepackage{booktabs}
\usepackage{url}

\usepackage{balance}
\usepackage{breakurl}
\usepackage{amsmath}
\IEEEoverridecommandlockouts
\IEEEpubid{\makebox[\columnwidth]{978-1-6654-9952-1/22/\$31.00~\copyright{}2022 IEEE \hfill} \hspace{\columnsep}\makebox[\columnwidth]{ }}
\begin{document}
\title{ETHER\textbf{LED}: Sending Covert Morse Signals from Air-Gapped Devices via Network Card (NIC) LEDs}


\author{\IEEEauthorblockN{Mordechai Guri}
	\IEEEauthorblockA{Ben-Gurion University of the Negev, Israel\\
		Department of Software and Information Systems Engineering\\ Cyber-Security Research Center \\
		Email: gurim@post.bgu.ac.il \\ Air-gap research page: http://www.covertchannels.com}}

\maketitle              % typeset the header of the contribution
%
%Highly secure devices are often isolated from the Internet or other public networks due to the confidential information they process. This level of isolation is referred to as an 'air-gap .' With air-gap isolation, even if the device is compromised, an attacker cannot leak data out of the device because of the lack of Internet connection. 
%
%In this paper, we present a new technique named ETHERLED, allowing attackers to leak data from air-gapped networked devices such as PCs, printers, network cameras, embedded controllers, and IoT devices. Networked devices have an integrated network interface controller (NIC) that includes status and activity indicator LEDs. We show that malware installed on the device can control the status LEDs by blinking and alternating colors, using documented methods or undocumented firmware commands. Information can be encoded via simple encoding such as Morse code and modulated over these optical signals. An attacker with a line-of-sight to the status LEDs can intercept and decode these signals from tens to a hundred meters away. We examine the attack vector and present design and implementation. We show an evaluation and discuss defensive and preventive countermeasures for this exfiltration attack.

\begin{abstract}
Highly secure devices are often isolated from the Internet or other public networks due to the confidential information they process. This level of isolation is referred to as an 'air-gap .' 

In this paper, we present a new technique named ETHERLED, allowing attackers to leak data from air-gapped networked devices such as PCs, printers, network cameras, embedded controllers, and servers. Networked devices have an integrated network interface controller (NIC) that includes status and activity indicator LEDs. We show that malware installed on the device can control the status LEDs by blinking and alternating colors, using documented methods or undocumented firmware commands. Information can be encoded via simple encoding such as Morse code and modulated over these optical signals. An attacker can intercept and decode these signals from tens to hundreds of meters away. We show an evaluation and discuss defensive and preventive countermeasures for this exfiltration attack.

\begin{IEEEkeywords}
	air-gap, exfiltration, covert channel, optical, Ethernet.
\end{IEEEkeywords}
%\keywords{air-gap  \and exfiltration \and covert channel}
\end{abstract}
%
%But some critical systems aren’t exposed to the public internet and sit, apparently safely, in an isolated environment, air gapped from the 
%
\section{Introduction}
The term `air-gap' in network security refers to a policy where computers, networks, or devices are not exposed to the public Internet. This isolation level is achieved by physically isolating the critical systems from the external world networks, especially the Internet. The air-gap policy is widely used in military and defense systems, critical infrastructure, governmental agencies, finance organizations, and other industries \cite{AirGappe12:online}. The air-gap isolation is maintained by enforcing strict policies in the organization. These policies include forbidding external devices such as Wi-Fi and Bluetooth adapters in secure environments. In order to protect the air-gapped network, intrusion detection and prevention systems may be used to eliminate any intentional or accidental security breaches. Traditionally, air-gap isolation was used inside a controlled networking environment called a SCIF (Sensitive Compartmented Information Facility) \cite{Implemen10:online}. A known example of an air-gapped network is the Joint Worldwide Intelligence Communications System, a classified network belonging to the United States Defense Intelligence Agency. 
 
\subsection{Air-gap Breaches}
In the past decade, it has been shown that even air-gapped networks are not immune to breaches. As noted by ESET research, in the first half of 2020 alone, four previously unknown malicious frameworks targeting air-gapped networks were found \cite{dorais2021jumping}. Adversaries could use complex attack vectors to hack air-gapped networks, such as supply chain attacks and malicious insiders. Attackers can penetrate an air-gapped network using these methods while bypassing defense measures, including firewalls, antivirus programs, intrusion detection, and prevention systems. For example, a classified network of the United States military was compromised by a computer worm named Agent.Btz \cite{AgentBTZ65:online} when a foreign intelligence agency supplied infected thumb drives to the target military base. The malicious thumb drive was put into a USB port of a laptop computer that was attached to United States Central Command. Stuxnet, ProjectSauron, and other APTs are examples of an air-gap breach that has also been reported in the past \cite{dorais2021jumping,guri2021power}.

\subsection{Leaking Information from Air-gapped Facilities}
With malicious code running in the air-gapped network, the attacker may want to retrieve sensitive information from the compromised network. For example, an attacker may wish to leak text files, encryption keys, and keylogging data. While the infection of air-gapped systems has been shown feasible, the exfiltration of data from off-line, disconnected systems is a much more challenging task. In order to leak the information despite the lack of Internet connectivity, the attack may resort to special communication techniques called air-gapped covert channels. Classic air-gap covert channels use various types of emanation and radiation from the target systems. Exploiting electromagnetic radiation (EMR) has been the most studied covert channel for twenty years. In this method, malware exploits the electromagnetic emission from various hardware components such as communication cables, processing units, and other hardware peripherals for data exfiltration. In recent years, sound waves, magnetic fields, and heat emissions have also been proposed as covert channels. 
At the optical domain, leaking information via the keyboard LEDs, hard drive LEDs, and screen power brightness \cite{guri2019brightness} was also proposed. In these methods, binary data is encoded over the activities of the LEDs and recorded by remote cameras. Most of the methods are not considered entirely covert and can be detected by users who notice irregular LED blinking patterns. 

\subsection{Our Contribution}
In this paper, we propose a new optical covert channel for the exfiltration of information from air-gapped systems. The attack is relevant to many devices in the modern IT environments which are shipped with integrated Ethernet cards, including:

\begin{enumerate}
	\item Personal workstations (PC), laptops, and servers
	\item Networked Attached Storage (NAS)
	\item Internet of Things (IoT) devices
	\item Printers and scanners
	\item TVs and LCD screens
	\item Embedded systems and controllers
	\item Security and surveillance cameras
\end{enumerate} 

We demonstrate how attackers can control the status network interface LEDs and encode data over it in a stealth way. We show that a local or remote video receiver with a line-of-sight with the device can record the LED activity and send it to the attacker for decoding. We discuss the related work in Section \ref{sec:related}, present the adversarial attack model in Section \ref{sec:adv}, and examine the modulation techniques and different type of receivers in Section \ref{sec:trans}. We evaluate the proposed covert channel in Section \ref{sec:eval}, and present countermeasures in Section \ref{sec:cnt}.   

\section{Related Work}
\label{sec:related}
In computer security, a covert channel is an attack that enables transferring information between two entities (e.g., computers) that are not allowed to communicate. Over the years, many network protocols have been studied in the context of covert communication, including IP, TCP and UDP, HTTP, SMTP, DNS, and others \cite{mazurczyk2016information}. It is also possible to encode information in packet timing, and image data \cite{wendzel2014hidden}. Unlike the traditional covert channels, air-gap covert channels focus on data leakage from offline computers, where there is no Internet connectivity. Air-gap covert channels are categorized into electromagnetic, acoustic, thermal, and optical channels.

Electromagnetic covert channels involve the usage of electromagnetic (or magnetic) emissions generated by the target device to carry information. Kuhn presented an attack that transmits data from a video cable via radio frequencies \cite{kuhn1998soft}. Guri et al. introduced AirHopper \cite{guri2014airhopper} and Air-Fi \cite{guri2020air} attacks aimed bridging the air-gap between computers and a nearby smartphones by exploiting signals at various frequency bands. Other types of magnetic and electric covert channels have been discussed in recent years \cite{burton2021private}. Bauer et al. discussed the exploitation of anti-EMI features of processors for covert communication \cite{bauer2016information}. Camurati presented a covert channel that affects mixed-signal chips, where electromagnetic leakage from digital logic is modulated with the radio carrier \cite{camurati2018screaming}.
Another type of air-gap covert channel relies on the acoustic medium for transmission. Researchers introduced a method called acoustical mesh networks, which enables the transmission of data between computers via high-frequency acoustic sound \cite{Hanspach2013}. Guri et al. presented Fansmitter \cite{guri2020fansmitter} and PowerSupply \cite{guri2021power}, methods enabling exfiltration of data via sound waves, even when the computers are not equipped with speakers or audio hardware. 
Takymchuk et al. presented a temperature based covert channel for FPGA systems \cite{iakymchuk2011temperature}. Other types of attacks control the device heat emissions to maintain a so-called thermal covert channel \cite{guri2015bitwhisper}. 

\subsection{Optical}
Leaking data from air-gap via the optical medium has also been proposed over the years. Twenty years ago, Loughry discussed the threat of information leakage via optical emanations via network devices and keyboard LEDs \cite{loughry2002information}. Loughry also discussed the optical injection attacks using LED status indicators \cite{loughry2019oops} and explores the threat of optical tempest in general \cite{loughry2018optical}. More recently, Guri presented the attack model involving information leakage through signals sent from the USB keyboard \cite{guri2019ctrl} and hard disk drives (HDD) LEDs \cite{Guri2017}. Lopes presented a method for leaking data using infrared LEDs shipped on storage devices \cite{lopes2017platform}. Note that their approach requires the attacker to deploy the compromised hardware in the organization. Nassi et al. showed how an attacker could exploit the light sensitivity of a scanner and use external light sources to deliver commands to malware in the network \cite{nassi2018xerox}. VisiSploit is another type of optical covert channel in which data is leaked through an image concealed on an LCD screen \cite{guri2019optical}. With this method, the invisible QR image that is shown on the computer screen is obtained by a remote camera and is then decoded using image processing. Weis discussed Aa QR-Code based optical covert channel in an air-gapped infrastructure \cite{weise2022qr}. Some works proposed to project invisible images on a modified LCD screen for data exfiltration. Such methods are less practical in a real attack model since they require hardware modification of the screen hardware.

This study uses the network devices Ethernet LEDs for data leakage, a threat that has not been studied before, theoretically or technically. We examine the opportunities and limitations of this technique on different types of transmission techniques and examine different types of cameras in the attack model. 

\begin{figure*}
	\centering
	\includegraphics[width=\linewidth]{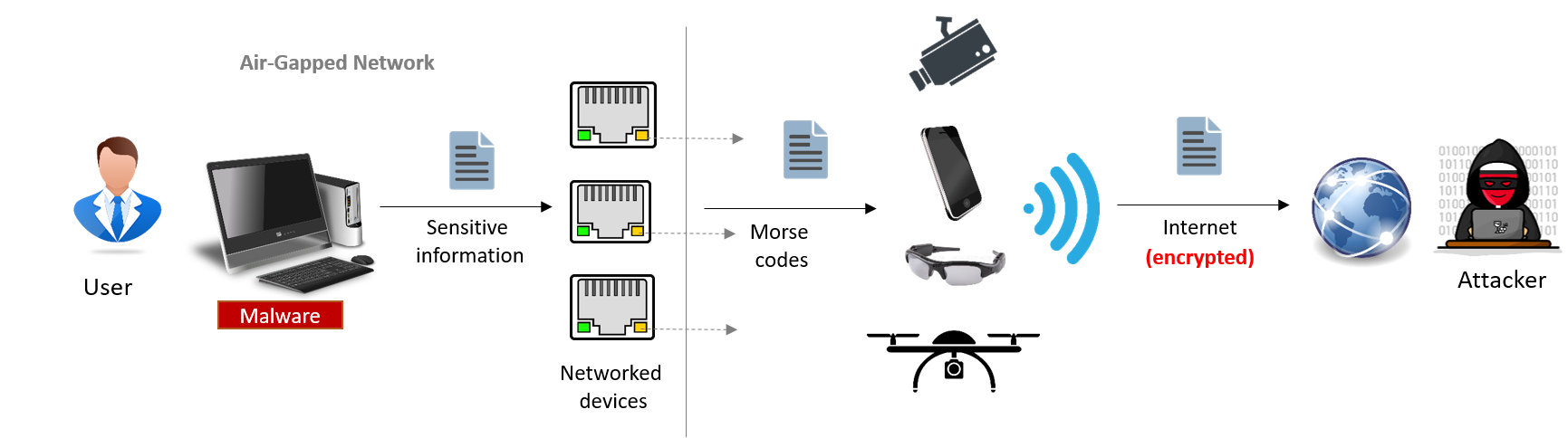}
	\caption{Adversarial attack model. An APT collects sensitive information and encodes on over NIC LEDs control. The data could be recorded by different types of cameras and transmitted to the attacker via the Internet.}
	\label{fig:attack}
\end{figure*}

\section{Adversarial Attack Model}
\label{sec:adv}
The term advanced persistent threat (APT) refers to an attack campaign in which an adversary maintains an illicit presence on a target network in order to access highly sensitive data \cite{virvilis2013big}. The attack model of ETHERLED is bifurcated and contains transmitter and receiver components. At the initial phase of the attack, the adversary must execute malicious code within the targeted system to enable control of the NIC LEDs. The target might be an isolated or highly secure computer or other networked devices such as embedded systems, printers, cameras, and any device with a wired network interface. 
The infection of a device can be achieved via supply chain attacks, social engineering techniques, or the use of hardware with installed software or firmware. Hijacking secured and even air-gapped facilities is a type of attack that has been demonstrated in recent years. In 2020, researchers in a security firm found an Advanced Persistent Threat (APT) dubbed USBCulprit
seems to be designed to breach air-gapped networks \cite{guri2021usbculprit}. Other air-gapped targeting APTs include Agent.Btz \cite{AgentBTZ65:online}, Stuxnet \cite{kushner2013real}, Tick and Ramsay \cite{NewRamsa19:online} attacks.  

\subsection{Data Exfiltration}
At the second phase of the attack, the malicious code collects data from the compromised network. The data can be textual (e.g., user names, passwords, key-loggings) or binary (e.g., encryption keys, biometric information). After collecting the information, the malware starts the exfiltration phase. In our case, the attack uses the optical covert channel for the data exfiltration. The malicious code uses the network card's status LEDs to encode the textual or binary information in stealth and covert ways, such as optical data modulations and Morse codes.    

\subsection{Data Reception}
The attacker receives the optical signals via cameras with a line of sight with the compromised device NIC LEDs. Several types of equipment can be used in this attack model. The attack may attack video surveillance, closed-circuit TV, or IP camera positioned in a location with a line of sight with the transmitting computer. An attacker may also use a hidden camera that has a line of sight to the transmitting device. In March 2021, a group of hackers compromised large group security cameras via a Silicon Valley startup. Due to the CheckPoint security report from 2021, over 150,000 IP cameras inside public organizations such as hospitals, police departments, and schools were compromised by hackers \cite{HackersG3:online}.
Another potential receiver is a drone with a high-resolution camera positioned, so it has a line of sight with the target device. Attackers can also use a malicious insider carrying a recording device (e.g., smartphone) that can position themself to have a line of sight with the transmitting device. These video cameras record the device's NIC LEDs. Then the attacker can then decode the signals and retrieve the encoded information. The list of potential receivers and the corresponding attack vectors is presented in Table \ref{tab:receivers}. 

% \usepackage{booktabs}
\begin{table}[]
	\caption{The list of potential receivers and the corresponding attack vectors}
	\label{tab:receivers}
	\begin{tabular}{@{}ll@{}}
		\toprule
		Receiver                     & Attack vector                         \\ \midrule
		Surveillance/security camera & Remote attack; exploitation            \\
		Hidden camera                & Supply chain attack; malicious insider \\
		Drone camera                 & Outsider attack                       \\
		Recording camera; smartphone & Malicious insider/visitor             \\ \bottomrule
	\end{tabular}
\end{table}

Figure \ref{fig:attack} shows the adversarial attack model. An APT collect sensitive information and encode on over NIC LEDs control. The data is recorded by different types of cameras and transmitted to the attacker via the Internet.

\section{Data Transmission and Reception}
\label{sec:trans}
This section describes the data transmission and discusses signal generation, modulation, and information encoding. Note that the topic of LED-based communication has been widely studied in the context of optical communication channels. The relevant literature discusses topics such as the type of optical receivers, effective distance, and environmental conditions in the general case \cite{jovicic2013visible}. This paper limits the discussion to NIC LEDs, their characteristics, and their relevancy to the attack model. 

\subsection{Signal Generation}
Network adapters are usually implemented as extension cards or hosted (integrated) in the PC motherboards and device system-on-chip. Common Ethernet port has two LEDs: the link status LED and the activity LED. The link status LED is a two-color LED, typically green and amber, while the activity LED is a one-color LED, usually green. Note that the behavior of the LEDs and their colors may be different from system to system. Typically the status LED color indicates the current link speed, e.g., amber for 1GB, green for 100Mb, and off for 10Mb connection. The activity LED blinks if there is a network activity in this port and is turned off if no link is established. Table \ref{tab:leds} Summarized typical configuration and coloring scheme of two NIC LEDs.

% \usepackage{booktabs}
\begin{table}[]
	\caption {The NIC LEDs used for the covert channel}
	\label {tab:leds}
	\begin{tabular}{@{}lllll@{}}
		\toprule
		LED      & Colors (optional) & Steady  & Blinking & Description                      \\ \midrule
		Link     & Green/Amber/Blue  & Enabled & Enabled  & Link status \& speed\\
		Activity & Green/Yellow/Red  & Enabled & Enabled  & Port activity         \\ \bottomrule
	\end{tabular}
\end{table}

\begin{figure}
	\centering
	\includegraphics[width=0.9\linewidth]{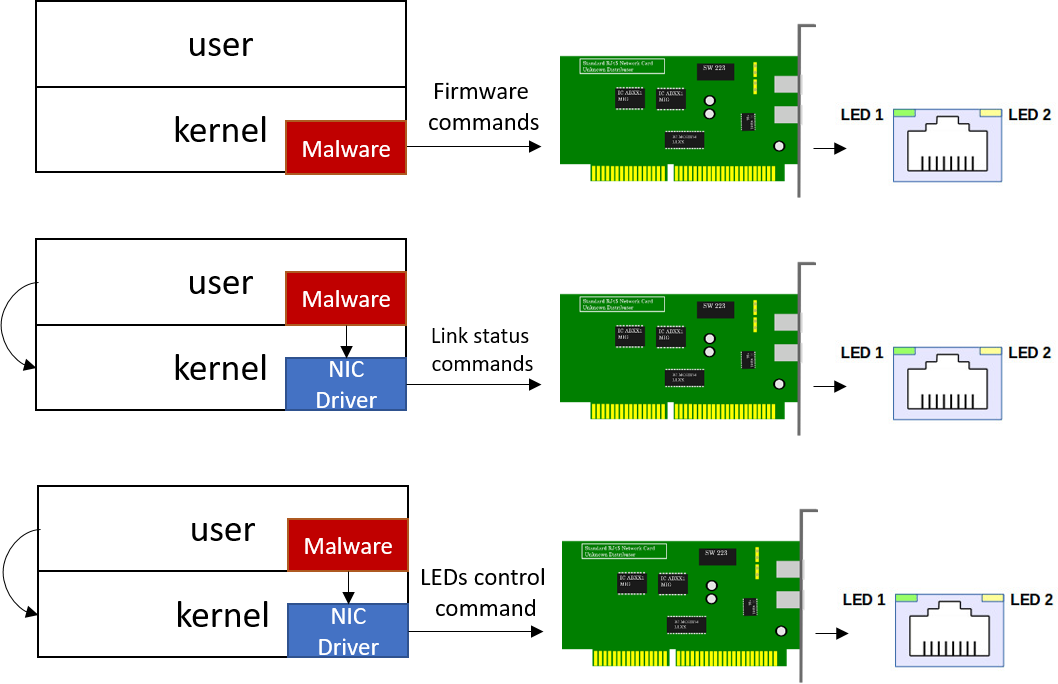}
	\caption{The three LEDs control methods.}
	\label{fig:arch}
\end{figure}

There are three main methods that malware can control the NIC LEDs (Figure \ref{fig:arch}).

\subsubsection{Driver/firmware control}

\begin{figure}[]
%	\centering
	\caption{Example of the embedded Ethernet LEDs control via GPIOs}
	\label{fig:code1}
	\frame{\includegraphics[width=1\linewidth]{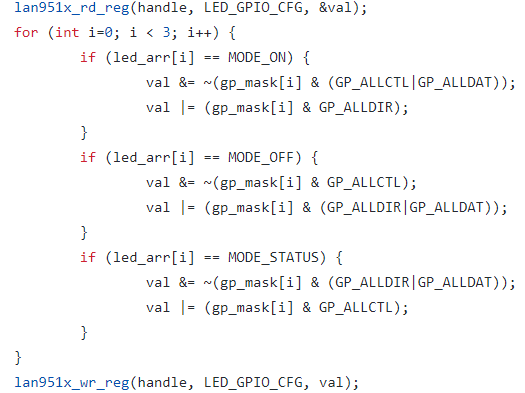}}

\end{figure}
%
%\begin{figure}[]
%	%	\centering
%	\caption{Embedded Ethernet LED control via the internal USB controller.}
%	\label{fig:code2}
%	\frame{\includegraphics[width=1\linewidth]{emb.png}}
%	%\includegraphics[width=0.9\linewidth]{arch2.png}
%	
%\end{figure}

In this method, the LED controlling code runs as a kernel driver or within the NIC firmware. Changing the LED state/color requires direct access to low-level registers or special non-volatile memory (NVM) addresses. This method enables the highest degree of control over the LEDs but is very hardware-specific and mostly undocumented. For example, the guide in \cite{Questtod45:online} discusses how to control the Ethernet LEDs in Intel NUC PC D34010WY. It can be done from a kernel driver (e.g., the \texttt{e1000\_led\_off\_pchlan} function) or by writing to specific addresses in flash memory (word 0x18), which holds LEDs configuration. For embedded controllers, the control on the NIC is typically performed via internal BUS or USB interfaces. For example, the sample code in \cite{lan951xl77:online} programs the corresponding LED register (LED\_GPIO\_CFG) for LAN951X Ethernet controllers via USB commands (Fig. \ref{fig:code1}).

\subsubsection{Link status control}
In this method, only the status LED can be controlled. The malicious code can intentionally change the link speed, which in turn causes the network adapter to change the status LED. For example, setting the link speed to 10Mb, 100Mb and 1Gb will set the status LED to off, green, and amber, respectively. Selecting the link speed can be done by interacting with the NIC driver. For example, the \texttt{ethtool} command-line tool in Linux enables to change of the link speed of the Ethernet controller \cite{ethtool848:online}. The same is possible in the Windows OS via the \texttt{netsh} command \cite{NetshCom27:online}. Note that setting the link speed requires root/admin privileges in both the Linux and Windows OSs. Technically, the link speed is determined through the \textit{autonegotiation} procedure. In this procedure, which occurs in the physical layer, the connected devices share their capabilities regarding supported parameters such as transmission rate, half/full duplex, etc. The link speed of the network card (NIC) can be determined from the computer's OS. 

\subsubsection{User LEDs control}
In this method, the user directly turns the status LEDs on and off by enabling and disabling the Ethernet interface using API or tools such as the \texttt{ethtool} or \texttt{eth}. The user directly turns the status LEDs on and off by enabling and disabling the Ethernet interface using API or tools such as the \texttt{ethtool} or \texttt{eth}. Another technique to blink the status LED is using the 'test' or 'identify' functionality, enabling the operator to identify the adapter by visual indication. These operations can be triggered programmatically or via low-level tools such as \texttt{ethtool}. 

Table \ref{tab:tech} presents the three LED control techniques and their corresponding capabilities for the covert channel.

% \usepackage{booktabs}
\begin{table}[]
	\caption {LED control techniques and t capabilities}
	\label {tab:tech}
	\begin{tabular}{@{}llll@{}}
		\toprule
		Technique               & LEDs control  & Color control & Blink control \\ \midrule
		Driver/firmware control & Link/Activity & Yes           & Yes           \\
		Link status control     & Link          & No            & Yes           \\
		User LEDs control       & Link/Activity & Yes           & Yes (limited) \\ \bottomrule
	\end{tabular}
\end{table}

\subsection{Data Modulation}
Using the signal generation techniques described above, we implemented three different types of data modulation; (1) on-off keying (OOK), (2) blink frequency, and (3) color modulation.

\subsubsection{On-Off keying}
On-Off keying is an basic form of the more general amplitude-shift keying (ASK) modulation. The lack of a signal for a certain time duration encodes a  zero ('0'), while its presence for a time duration encodes a one ('1'). In our case, LED is turned off for duration of T$_{0}$ encodes '0' and turned on for a duration T$_{1}$ encodes '1.' Note that in the simple case T$_{0}$ = T$_{1}$. In its simple implementation, this scheme uses a single LED to modulate data.
 
\subsubsection{Blinks frequency}
Frequency-shift keying (FSK) is a modulation scheme in which digital information is modulated through frequency changes in a carrier signal. In binary frequency-shift keying (B-FSK), only two frequencies are used. In our case, LED blinks in a frequency of F$_{0}$ encodes '0' LED blinks in a frequency of F$_{1}$ encodes '1'. We make a separation between two sequential bits by setting the LED in the off state for time interval F$_{d}$. This scheme uses a single LED to modulate data in its basic form. 

\subsubsection{Color modulation}
\begin{figure}
  		\centering
  		\includegraphics[width=0.9\linewidth]{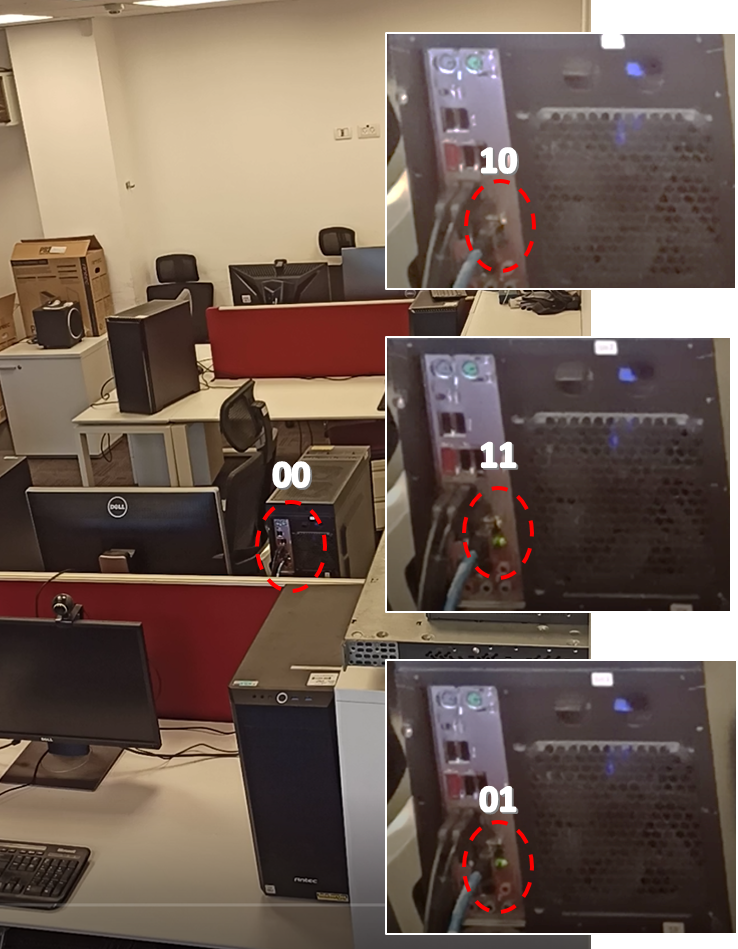}
  		\caption{Transmission of the sequence \texttt{00101101} (representing the 0x2d byte) using color modulation.}
  		\label{fig:il2}
  	\end{figure}

In the color modulation scheme, different LED colors are used to modulate the information. This can be seen as a form of frequency modulation at the carrier level since different colors have a different wavelength from the optical perspective. In our case, the link LED blinks encodes '0', and the activity LED blinks encodes '1'. Another version of this modulation uses the same LED with different colors, e.g., by alternating the status LED color between green and amber. We make a separation between two sequential bits by setting the LED in the off state for time interval T$_{d}$. In this form, the scheme uses two LEDs to modulate data. Figure \ref{fig:il2} shows the transmission of the sequence \texttt{00101101} (0x2d) using color modulation, as recorded by a video camera in the room.

\subsection{Data Frames}
\begin{figure}[!h]
	\centering
	\includegraphics[width=0.8\linewidth]{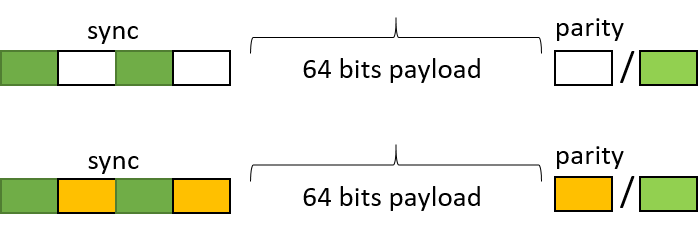}
	\caption{A data frame with a single LED (top) and two LEDs (bottom) modulations.}
	\label{fig:frame}
\end{figure}

We transmit the data in defined frames depicted in Figure \ref{fig:frame}. Each frame begins with a sequence \texttt{`1010'} followed by a payload of 64 bits. Note that the encoding of `0' and `1' depends on the modulation used (e.g., single color or two colors). The frame ends with a parity bit for error detection.

\subsection{Morse Codes Encoding}
We also implemented a Morse Codes transmitter to transmit textual information. Morse code is an encoding scheme that allows sending messages via sequences of short and long signals, represented as dots and dashes, respectively. Since the Morse code requires control over a single LED, it enables the simple encoding of textual information via the network status LED. The Morse codes consist of short signals called dots, and long signals called dash. A dot is an elementary time unit, and a dash is three times as long as a dot. The separation between dots and dash within a character is a one-time unit. The separation between characters is three time-units, and between letters is seven-time units. In the case of NIC LEDs, the encoding is done on a single color LED, usually the green color. The time units may vary depending on the network cards. We used a dot time duration of 100 ms in our experiments and a dash time duration of 300 ms. For driver/firmware control, we used a dot time duration of 10 ms and a dash time duration of 30 ms. The Morse codes encoding parameters with the status LED is listed in Tables \ref{tab:morse2},\ref{tab:morse2a}.

\begin{table}[h]
	\centering
	\caption{Morse Codes Encoding with the status LED}
	\label{tab:morse2}
	\begin{tabular}{@{}lll@{}}
		\toprule
		Symbol                     & Status  & Duration       \\ \midrule
		Dot                        & LED on  & 100 ms         \\
		Dash                       & LED on  & 300 ms         \\
		Spaces (short/medium/long) & LED off & 100/300/700 ms \\ \bottomrule
	\end{tabular}
\end{table}

\begin{table}[h]
	\centering
	\caption{Morse Codes Encoding with the status LED (driver/firmware)}
	\label{tab:morse2a}
	\begin{tabular}{@{}lll@{}}
		\toprule
		Symbol                     & Status  & Duration       \\ \midrule
		Dot                        & LED on  & 10 ms         \\
		Dash                       & LED on  & 30 ms         \\
		Spaces (short/medium/long) & LED off & 10/30/70 ms \\ \bottomrule
	\end{tabular}
\end{table}
 
% \usepackage{booktabs}

\section{Evaluation}
\label{sec:eval}
In this section, we evaluate the optical covert channel. Our evaluation focuses on the optical characteristics of NIC LEDs and the transmission rate. Our experiments adapt the approach commonly used in visible light communication (VLC), which assumes a line of sight between the light source and the camera. Note that other work focused on retrieving singles where the camera has no line-of-sight with the transmitting source \cite{jovicic2013visible}. We implemented a prototype of the described modulation schemes and Morse code transmitter on different PCs and laptops running Ubuntu OS in Python scripting language. We also used the command-line utilities such as the \texttt{ethtool} and \texttt{netsh}.

\subsection{Camera Receivers}
After retrieving the recorded video, the attacker has to perform the following video processing tasks; (1) detection of the location of each transmitting LED, (2) frame by frame processing to identify the LED status per frame, and (3) binary data is decoded based on the encoding scheme.

The main factor in determining the bit rate for video cameras is the number of frames per second. Our experiments identified three frames per bit as the optimal setting to decode the LED transmission successfully. In the context of the attack model, it means that even devices with low frame rates, such as security cameras, can be used for data recording. Table \ref{tab:distance} shows the effective distances with different types of receivers. Our evaluated video cameras include a high-end surveillance camera, entry-level DSLR camera, extreme high-resolution camera, HD webcam, smartphone camera, and telescope camera. As can be seen, the surveillance cameras and smartphones can reach lengths of 0 - 30 meters, while extreme cameras can reach up to 50 meters, depending on the resolution. Telescope and special focus lenses can extend the range to a few hundred meters \cite{jovicic2013visible}. 

% 
%
%\begin{figure}
%	\centering
%	\includegraphics[width=0.9\linewidth]{il4.png}
%	%\includegraphics[width=0.9\linewidth]{arch2.png}
%	\caption{Transmission of the sequence \texttt{00101101} (representing the 0x2d byte) using color modulation.}
%	\label{fig:il2}
%\end{figure}

% \usepackage{booktabs}
\begin{table}[]
	\caption {The effective distance with different receivers}
	\label {tab:distance}
	\begin{tabular}{@{}lll@{}}
		\toprule
		Receiver                    & Type                           & Distance \\ \midrule
		High-end surveillance camera & SNCEB600 Network 720p/30fps  & 30 m     \\
		Entry-level DSLR camera     & Nikon D7100 1920x1080 (video)  & 20 m     \\
		Extreme camera              & GoPro                          & 50 m     \\
		Webcam (HD)                 & Microsoft LifeCam              & 10 m     \\
		Smartphone camera           & Samsung Galaxy                 & 30 m     \\
		Focus lenses / telescope    & Various                        & 100+ m    \\ \bottomrule
	\end{tabular}
\end{table}

% \usepackage{booktabs}
\begin{table*}[]
	\caption {The effective bitrates with different modulations}
	\label {tab:bitrate}
	\begin{tabular}{@{}llll@{}}
		\toprule
		Technique               & Modulations /                & Max bit rate & Relevant information                                                                                \\ \midrule
		Driver/firmware control & OOK, Blink-frequency, colors & 100 bit/sec  & Text files, Username/passwords, encryption keys (e.g, RSA 4096), PIN codes                     \\
		Link status control     & OOK, Blink-frequency         & 1 bit/sec    & Username/passwords, credentials, encryption keys (e.g, AES 256), PIN codes                     \\
		User LEDs control       & OOK, Blink-frequency, colors & 2 bit/sec    & Keylogging, username/passwords, credentials, encryption keys (e.g, AES 256)  \\ \bottomrule
	\end{tabular}
\end{table*}

\subsection{Bit Rates}
Table \ref{tab:bitrate} shows the bit rates when various LED control techniques and modulations are used. For the user LED control, the effective bitrate is between 1 bit/sec with link status control to 2 bit/sec. The most relevant information to leak using this bit rate is credentials, username and password, encryptions keys (e.g., RSA 4096), and key-loggings. With the low-level driver/firmware control, we reached bitrates of 100 bit/sec, enabling the exfiltration of data such as text files. It is important to note that the devices with Ethernet cards are usually stationary and positioned in the same place. Maintaining a covert channel for an extended period is feasible since the line-of-sight with the target device is stable. 

\subsubsection{Morse codes}
Morse code speed is specified in words per minute (WPM). For the Morse codes bandwidth evaluation, the word PARIS is used as standard of the typical word in English plain text. The WPM is calculated by the formula 

\begin{equation}
	Speed (WPM) = 60 / (50 * Dotlength (sec))
\end{equation}

In our case, with a Dot length of 0.3 sec, the Speed (WPM) = 4 (four words per second). With a Dot length of 0.03 sec (driver/firmware control), the Speed (WPM) = 40 words per second.

\subsection{Data Transmission}
Tables \ref{tab:data},\ref{tab:data2} present the time it takes to leak various types of information. As can be seen, a 100 bits password can be exfiltrated in 1 sec - 1.5 minutes, depending on the modulation used. Depending on the modulation used, a bit Bitcoin private key can be exfiltrated in approximately 2.5 sec - 4.2 minutes. Depending on the modulation used, the RSA encryption key of 4096 bits can be exfiltrated in 42 sec - 60 minutes. For textual keylogging, each keystroke can be leaked in almost real-time, and it's primarily relevant for fast modulations. 1KB of text (8000 bits) can be leaked with the driver/firmware control in 1.4 - 2.7 min.
% \usepackage{booktabs}
\begin{table}[]
	\caption{time it takes to leak various types of information}
	\label{tab:data}
	\begin{tabular}{@{}llcc@{}}
		\toprule
		Information         & Size      & \multicolumn{1}{l}{Single color} & \multicolumn{1}{l}{Two colors} \\ \midrule
		Passwords           & 100 bits  & 1.5 min                                      & 0.7 min                                \\
		Bitcoin private key & 256 bits  & 4.2 min                                      & 2.1 min                                \\
		PIN codes           & 64 bits   & 1 min                                        & 0.5 min                                \\
		RSA encryption keys & 4096 bits & $\sim$30 min                                 & $\sim$60 min                           \\
		Keylogging          & 5 bit/key & N/A                                          & 2 sec / key                            \\ \bottomrule
	\end{tabular}

\end{table} 

\begin{table}[]
	\caption{time it takes to leak various types of information (driver/firmware)}
	\label{tab:data2}
	\begin{tabular}{@{}llcc@{}}
		\toprule
		Information         & Size      & \multicolumn{1}{l}{Single color} & \multicolumn{1}{l}{Two colors} \\ \midrule
		Passwords           & 100 bits  & 2 sec                                       & 1 sec                                \\
		Bitcoin private key & 256 bits  & 5 sec                                      & 2.5 sec                                \\
		PIN codes           & 64 bits   & 1.2 sec                                        & 0.5 sec                                \\
		RSA encryption keys & 4096 bits & 1.4 min                                 & 42 sec                           \\
		Text files          & 1KB &  $\sim$2.7 min                                          & $\sim$ 1.4 min                          \\ \bottomrule
	\end{tabular}
\end{table}

\section{Countermeasures}
\label{sec:cnt}
TEMPEST attack, taken from the National Security Agency (NSA) jargon, loosely refers to the threat of data leakage from systems and devices through leaked emanations, including electromagnetic, acoustic, and optical. The defensive countermeasures for optical TEMPEST include the restriction of cameras and video recorders in areas with line-of-sight with the sensitive devices. However, many types of equipment, such as surveillance cameras, may be installed to monitor the sensitive areas, so they can not be banned entirely. It is important to note that the receiving camera is not necessarily a stationary one but might be a transitional, such as a camera carried on a spying drone. 
Nassi et al. introduced a method that can detect whether a specific spot (e.g., a room) is being video streamed by a drone camera \cite{nassi2019drones}. They maintained a periodic optical stimulus, which was reflected as a watermark in the wireless traffic. Particular types of opaque windows might be used to prevent the threat of remote cameras, drones, and optical leakage. However, this type of countermeasure doesn't protect against insider attacks or cameras within the building. 

Another physical countermeasure against the ETHERLED attack is to cover the status LEDs with black tape to block the optical emanation physically. It is also possible to disconnect the LEDs power input at the hardware level. The two above solutions are less practical since they affect the usability of the status LEDs, which are necessary in many cases. At the software level, it is possible to reprogram the NIC EEPROM settings to fixate the LEDs' colors and blinking frequency, preventing the modulation and encoding scheme described in this paper. A jamming approach adds noise to the modulated signals by invoking random LEDs indications. The status signals generated by the malicious code will get mixed up with random noise. Similar to the previous solutions, this type of software intervention degrades the usability of the status LEDs. Another possible countermeasure is to detect covert signaling patterns, encoding, and modulations. These solutions can be implemented internally in the OS or externally from another device. Kernel-level drivers monitor the command sent to the NICs and detect suspicious patterns in the internal solution, such as repetitive speed changes or direct LEDs commands. Note that any type of OS-level solution can be bypassed by advanced code in the kernel. Since modern Smart NICs (sNICs) are fully programmable, it is possible to offload the covert channel detection and prevention to the sNIC \cite{feng2020smartnic,liu2019offloading}. However, note that the attack presented in this paper is relevant to any type of NIC. Previous work presented different approaches for the detection of covert channels. Carrara presented a survey and taxonomy of detection and measurement of various covert channels  \cite{carrara2016survey}. Cabaj et al. discussed using a data-mining approach to detect network covert channel \cite{carrara2016survey}. Another type of detection and prevention countermeasures can be deployed in network devices. A traffic inspection is executed in this solution's local router or switch. The component detects and blocks suspicious activities such as frequent changes in the interface speed or abnormal traffic patterns. Note that this type of solution is considered trusted but requires cooperation from the network device manufacturers. An external solution includes a camera that monitors a specific device or even the entire room and detects anomalies in optical activities such as LED communication. A practical implementation of external monitoring would be complex because network device LEDs routinely blink due to regular traffic activity. Consequentially, this kind of detection solution would likely have high rates of false positives. Table \ref{tab:cnt} summarizes the countermeasures to the ETHERLED attack.

% \usepackage{booktabs}
\begin{table}[]
	\caption{Defensive and preventive countermeasures}
	\label{tab:cnt}
	\begin{tabular}{@{}ll@{}}
		\toprule
		Countermeasure                & Weaknesses                               \\ \midrule
		Camera zone restriction               & Cost and space                           \\
		OS monitoring/blocking        & Can be evaded; false positives           \\
		Smart NIC monitoring/blocking & Not applicable in all cases              \\
		Signal jamming                & Malicious insiders; supply-chain attacks \\
		Optical watermarking          & Specific to wireless cameras             \\
		Optical monitoring            & Cost; deployment; false positives        \\
		Network device detection      & Deployment; OEM involvement                         \\ \bottomrule
	\end{tabular}
\end{table}

\section{Conclusion}
This paper showed how an attacker could exfiltrate information from a wide range of highly secured, air-gapped devices such as PCs, printers, network cameras, IoT devices, and embedded controllers. These devices have an integrated network interface controller (NIC), including status and activity indicator LEDs. We showed that malware installed on the device could programmatically control the status LED by blinking or alternating its colors. We presented different levels of slow and fast control on the status LEDs via link-layer and firmware level commands. Textual and binary information such as passwords and encryption keys can be encoded and modulated over these signals. An attacker with a line-of-sight to the status LEDs can intercept and decode these signals. e.g., via remote drone or local surveillance camera. We also examined different types of cameras receivers, such as smartphone and security cameras. We discussed various data encoding and modulation schemes and proposed defensive countermeasures for this attack. 

\balance      
\bibliographystyle{plain}
\bibliography{../../AirGap,../../AirGapCases,NIC}

\begin{thebibliography}{10}

\bibitem{AgentBTZ65:online}
Agent.btz - wikipedia.
\newblock \url{https://en.wikipedia.org/wiki/Agent.BTZ}.
\newblock (Accessed on 09/05/2022).

\bibitem{AirGappe12:online}
Air gapped networks: A false sense of security? - sentinelone.
\newblock
  \url{https://www.sentinelone.com/blog/air-gapped-networks-a-false-sense-of-security/}.
\newblock (Accessed on 03/01/2022).

\bibitem{ethtool848:online}
ethtool(8) - linux man page.
\newblock \url{https://linux.die.net/man/8/ethtool}.
\newblock (Accessed on 03/04/2022).

\bibitem{HackersG3:online}
Hackers gained access to 150,000 ip cameras inside hospitals, police
  departments, prisons, schools, and companies like tesla \& equinox - check
  point software.
\newblock
  \url{https://blog.checkpoint.com/2021/03/26/hackers-gained-access-to-150000-ip-cameras-inside-hospitals-police-departments-prisons-schools-and-companies-like-tesla-equinox/}.
\newblock (Accessed on 03/04/2022).

\bibitem{Implemen10:online}
Implementing a modern-day air gap network — qrypt.
\newblock
  \url{https://www.qrypt.com/blog/2021/8/17/implementing-a-modern-day-air-gap-network-6d39b}.
\newblock (Accessed on 03/01/2022).

\bibitem{lan951xl77:online}
lan951x-led-ctl/lan951x-led-ctl.c at master · dumpsite/lan951x-led-ctl ·
  github.
\newblock
  \url{https://github.com/dumpsite/lan951x-led-ctl/blob/master/src/lan951x-led-ctl.c}.
\newblock (Accessed on 05/23/2022).

\bibitem{NetshCom27:online}
Netsh command syntax, contexts, and formatting | microsoft docs.
\newblock
  \url{https://docs.microsoft.com/en-us/windows-server/networking/technologies/netsh/netsh-contexts}.
\newblock (Accessed on 03/04/2022).

\bibitem{NewRamsa19:online}
New ramsay malware can steal sensitive documents from air-gapped networks |
  zdnet.
\newblock
  \url{https://www.zdnet.com/article/new-ramsay-malware-can-steal-sensitive-documents-from-air-gapped-networks/}.
\newblock (Accessed on 09/05/2022).

\bibitem{Questtod45:online}
Quest to disable lan leds of an intel nuc.
\newblock \url{https://pwmon.org/p/1900/quest-disable-lan-leds-intel-nuc/}.
\newblock (Accessed on 03/04/2022).

\bibitem{‘Tick’es80:online}
‘tick’ espionage group is likely trying to hop air gaps, researchers say.
\newblock
  \url{https://www.cyberscoop.com/tick-espionage-usb-air-gaps-palo-alto-networks/}.
\newblock (Accessed on 09/05/2022).

\bibitem{bauer2016information}
Johannes Bauer, Sebastian Schinzel, Felix Freiling, and Andreas Dewald.
\newblock Information leakage behind the curtain: Abusing anti-emi features for
  covert communication.
\newblock In {\em 2016 IEEE International Symposium on Hardware Oriented
  Security and Trust (HOST)}, pages 130--134. IEEE, 2016.

\bibitem{burton2021private}
Thomas Burton and Kasper Rasmussen.
\newblock Private data exfiltration from cyber-physical systems using channel
  state information.
\newblock In {\em Proceedings of the 20th Workshop on Workshop on Privacy in
  the Electronic Society}, pages 223--235, 2021.

\bibitem{camurati2018screaming}
Giovanni Camurati, Sebastian Poeplau, Marius Muench, Tom Hayes, and
  Aur{\'e}lien Francillon.
\newblock Screaming channels: When electromagnetic side channels meet radio
  transceivers.
\newblock In {\em Proceedings of the 2018 ACM SIGSAC Conference on Computer and
  Communications Security}, pages 163--177, 2018.

\bibitem{carrara2016survey}
Brent Carrara and Carlisle Adams.
\newblock A survey and taxonomy aimed at the detection and measurement of
  covert channels.
\newblock In {\em Proceedings of the 4th ACM Workshop on Information Hiding and
  Multimedia Security}, pages 115--126. ACM, 2016.

\bibitem{dorais2021jumping}
Alexis Dorais-Joncas and Facundo Mun{\~o}z.
\newblock Jumping the air gap.
\newblock 2021.

\bibitem{feng2020smartnic}
Yixiao Feng, Sourav Panda, Sameer~G Kulkarni, KK~Ramakrishnan, and Nick
  Duffield.
\newblock A smartnic-accelerated monitoring platform for in-band network
  telemetry.
\newblock In {\em 2020 IEEE International Symposium on Local and Metropolitan
  Area Networks (LANMAN}, pages 1--6. IEEE, 2020.

\bibitem{guri2019optical}
Mordechai Guri.
\newblock Optical air-gap exfiltration attack via invisible images.
\newblock {\em Journal of Information Security and Applications}, 46:222--230,
  2019.

\bibitem{guri2020air}
Mordechai Guri.
\newblock Exfiltrating data from air-gapped computers via vibrations.
\newblock {\em Future Generation Computer Systems}, 122:69--81, 2021.

\bibitem{guri2021power}
Mordechai Guri.
\newblock Power-supplay: Leaking sensitive data from air-gapped, audio-gapped
  systems by turning the power supplies into speakers.
\newblock {\em IEEE Transactions on Dependable and Secure Computing}, 2021.

\bibitem{guri2021usbculprit}
Mordechai Guri.
\newblock Usbculprit: Usb-borne air-gap malware.
\newblock In {\em European Interdisciplinary Cybersecurity Conference}, pages
  7--13, 2021.

\bibitem{guri2019brightness}
Mordechai Guri, Dima Bykhovsky, and Yuval Elovici.
\newblock Brightness: Leaking sensitive data from air-gapped workstations via
  screen brightness.
\newblock In {\em 2019 12th CMI Conference on Cybersecurity and Privacy (CMI)},
  pages 1--6. IEEE, 2019.

\bibitem{guri2014airhopper}
Mordechai Guri, Gabi Kedma, Assaf Kachlon, and Yuval Elovici.
\newblock Airhopper: Bridging the air-gap between isolated networks and mobile
  phones using radio frequencies.
\newblock In {\em Malicious and Unwanted Software: The Americas (MALWARE), 2014
  9th International Conference on}, pages 58--67. IEEE, 2014.

\bibitem{guri2015bitwhisper}
Mordechai Guri, Matan Monitz, Yisroel Mirski, and Yuval Elovici.
\newblock Bitwhisper: Covert signaling channel between air-gapped computers
  using thermal manipulations.
\newblock In {\em Computer Security Foundations Symposium (CSF), 2015 IEEE
  28th}, pages 276--289. IEEE, 2015.

\bibitem{guri2020fansmitter}
Mordechai Guri, Yosef Solewicz, and Yuval Elovici.
\newblock Fansmitter: Acoustic data exfiltration from air-gapped computers via
  fans noise.
\newblock {\em Computers \& Security}, page 101721, 2020.

\bibitem{guri2019ctrl}
Mordechai Guri, Boris Zadov, Dima Bykhovsky, and Yuval Elovici.
\newblock Ctrl-alt-led: Leaking data from air-gapped computers via keyboard
  leds.
\newblock In {\em 2019 IEEE 43rd Annual Computer Software and Applications
  Conference (COMPSAC)}, volume~1, pages 801--810. IEEE, 2019.

\bibitem{Guri2017}
Mordechai Guri, Boris Zadov, and Yuval Elovici.
\newblock {\em LED-it-GO: Leaking (A Lot of) Data from Air-Gapped Computers via
  the (Small) Hard Drive LED}, pages 161--184.
\newblock Springer International Publishing, Cham, 2017.

\bibitem{Hanspach2013}
Michael Hanspach and Michael Goetz.
\newblock On covert acoustical mesh networks in air.
\newblock {\em Journal of Communications}, 8(11):758--767, November 2013.

\bibitem{iakymchuk2011temperature}
Taras Iakymchuk, Maciej Nikodem, and Kepa Krzysztof.
\newblock Temperature-based covert channel in fpga systems.
\newblock In {\em 6th International Workshop on Reconfigurable
  Communication-centric Systems-on-Chip (ReCoSoC)}, pages 1--7. IEEE, 2011.

\bibitem{jovicic2013visible}
Aleksandar Jovicic, Junyi Li, and Tom Richardson.
\newblock Visible light communication: opportunities, challenges and the path
  to market.
\newblock {\em IEEE communications magazine}, 51(12):26--32, 2013.

\bibitem{kuhn1998soft}
Markus~G Kuhn and Ross~J Anderson.
\newblock Soft tempest: Hidden data transmission using electromagnetic
  emanations.
\newblock In {\em Information hiding}, volume 1525, pages 124--142. Springer,
  1998.

\bibitem{kushner2013real}
David Kushner.
\newblock The real story of stuxnet.
\newblock {\em ieee Spectrum}, 3(50):48--53, 2013.

\bibitem{liu2019offloading}
Ming Liu, Tianyi Cui, Henry Schuh, Arvind Krishnamurthy, Simon Peter, and Karan
  Gupta.
\newblock Offloading distributed applications onto smartnics using ipipe.
\newblock In {\em Proceedings of the ACM Special Interest Group on Data
  Communication}, pages 318--333. 2019.

\bibitem{lopes2017platform}
Arthur~Costa Lopes and Diego~F Aranha.
\newblock Platform-agnostic low-intrusion optical data exfiltration.
\newblock In {\em ICISSP}, pages 474--480, 2017.

\bibitem{loughry2018optical}
Joe Loughry.
\newblock Optical tempest.
\newblock In {\em 2018 International Symposium on Electromagnetic Compatibility
  (EMC EUROPE)}, pages 172--177. IEEE, 2018.

\bibitem{loughry2019oops}
Joe Loughry.
\newblock (“oops! had the silly thing in reverse”)—optical injection
  attacks in through led status indicators.
\newblock In {\em 2019 International Symposium on Electromagnetic
  Compatibility-EMC EUROPE}, pages 376--382. IEEE, 2019.

\bibitem{loughry2002information}
Joe Loughry and David~A Umphress.
\newblock Information leakage from optical emanations.
\newblock {\em ACM Transactions on Information and System Security (TISSEC)},
  5(3):262--289, 2002.

\bibitem{mazurczyk2016information}
Wojciech Mazurczyk, Steffen Wendzel, Sebastian Zander, Amir Houmansadr, and
  Krzysztof Szczypiorski.
\newblock {\em Information hiding in communication networks: fundamentals,
  mechanisms, applications, and countermeasures}.
\newblock John Wiley \& Sons, 2016.

\bibitem{nassi2019drones}
Ben Nassi, Raz Ben-Netanel, Adi Shamir, and Yuval Elovici.
\newblock Drones' cryptanalysis-smashing cryptography with a flicker.
\newblock In {\em 2019 IEEE Symposium on Security and Privacy (SP)}, pages
  1397--1414. IEEE, 2019.

\bibitem{nassi2018xerox}
Ben Nassi, Adi Shamir, and Yuval Elovici.
\newblock Xerox day vulnerability.
\newblock {\em IEEE Transactions on Information Forensics and Security},
  14(2):415--430, 2018.

\bibitem{virvilis2013big}
Nikos Virvilis and Dimitris Gritzalis.
\newblock The big four-what we did wrong in advanced persistent threat
  detection?
\newblock In {\em 2013 international conference on availability, reliability
  and security}, pages 248--254. IEEE, 2013.

\bibitem{weise2022qr}
Martin Weise.
\newblock {\em A QR-Code Optical Covert Channel in an Air-Gapped Secure Data
  Infrastructure}.
\newblock PhD thesis, Wien, 2022.

\bibitem{wendzel2014hidden}
Steffen Wendzel, Wojciech Mazurczyk, Luca Caviglione, and Michael Meier.
\newblock Hidden and uncontrolled--on the emergence of network steganographic
  threats.
\newblock In {\em ISSE 2014 Securing Electronic Business Processes}, pages
  123--133. Springer, 2014.

\end{thebibliography}

\end{document}